\documentclass[nofootinbib,aps,11pt,preprintnumbers]{revtex4}
\usepackage{graphicx}
\usepackage{cancel}
\usepackage{amssymb}
\usepackage{textcomp}
\usepackage{amsmath}
\usepackage{bm}
\usepackage{times}
\usepackage{epsfig}
\usepackage{color}
\usepackage{graphics}
\usepackage{hyperref}
\usepackage{graphicx}
\usepackage{slashed}
\usepackage{hyperref}
\hypersetup{
    pdfnewwindow=true,     % links in new window
    colorlinks=true,       % false: boxed links; true: colored links
    linkcolor=black,       % color of internal links
    citecolor=blue,        % color of links to bibliography
    filecolor=blue,        % color of file links
    urlcolor=blue          % color of external links
}
\begin{document}
\title{\Large {\bf{Supersymmetry at the LHC and The Theory of R-parity}}}
\author{\large{Pavel Fileviez P\'erez}}
\affiliation{Particle and Astro-Particle Physics Division \\
Max Planck Institute for Nuclear Physics {\rm{(MPIK)}} \\
Saupfercheckweg 1, 69117 Heidelberg, Germany}
\author{\large{Sogee Spinner}}
\affiliation{International School for Advanced Studies {\rm{(SISSA)}} and INFN \\ Via Bonomea 265, 34136 Trieste, Italy}
\preprint{}
%%%%%%%%%%%%%%%%%%%%%%%%%%%%%%%%%%%%%%%%%%%
\begin{abstract}
We discuss the possible signatures at the Large Hadron Collider in models where R-parity is spontaneously broken.
In the context of the minimal gauge theory for R-parity, we investigate signals with multileptons which can provide an unique 
test of this theory. The possible impact of these ideas for the search of supersymmetry at the Large Hadron Collider 
is discussed. We also discuss the constraints coming from cosmology due to the existence of two light sterile neutrinos in the theory.
\end{abstract}
\maketitle
%\tableofcontents
%
%%%%%%%%%%%%%%%%%%%%%%%
\section{Introduction}
%%%%%%%%%%%%%%%%%%%%%%%
Despite the current lack of evidence for supersymmetry, the minimal supersymmetric standard model (MSSM) continues to be an appealing candidate for physics beyond the standard model (BSM) and tops the list of BSM searches for the CMS and ATLAS collaborations at the Large Hadron Collider (LHC). From the experimental searches so far, we have learned the following:

\begin{itemize}

\item The colored superpartners of the first and second generation, as well as the gluino, should be heavier than a TeV in order to avoid LHC bounds. Stop and sbottom mass bounds are weaker and therefore might be accessible during early runs of the 14 TeV LHC.

\item However, heavy stops and/or large left-right stop mixing are needed to accommodate the measured Higgs mass of around 125 GeV in the MSSM. Alternatively, it is possible to reduce the constraints on the stop sector in extensions of the MSSM through new F-terms, such as in the next to minimal supersymmetric standard model (NMSSM) or new D-terms.

\item If R-parity is broken it might be possible to avoid some of the LHC missing energy search bounds.

\end{itemize}
Furthermore, the fate of R-parity in the MSSM plays a crucial role in the possible discovery of supersymmetry at the LHC.
Therefore, it is crucial to understand all of the possible MSSM signatures with and without R-parity violation (RPV). This is an enormous task 
since the supersymmetric spectrum is not predicted in the context of the MSSM 
and there are many different decay channels leading to different signatures. In addition, if R-parity is violated, there is no prediction for which of the forty eight R-parity violating parameters in the superpotential are phenomenologically relevant or their relevant sizes.

This latter issue related to RPV	 is addressed in a model recently proposed by us: the simplest 
gauge theory for the fate of R-parity~\cite{FileviezPerez:2008sx, PRL}. In addition to predicting the spontaneous violation of R-parity, this theory further predicts that the R-parity violating terms must be bilinear and lepton number violating. Furthermore, since these couplings contribute to tree-level neutrino masses, their size is such that they can only affect the decay of the lightest supersymmetric particle (LSP) \footnote{Or the next to lightest supersymmetric particle (NLSP) in the case of a gravitino LSP.} leaving all other SUSY collider phenomenology unchanged. Therefore, one can say that this theory provides a concrete guide for R-parity violating phenomenology at the LHC. Based on $B-L$ gauge symmetry~\cite{PRL}, this theory predicts the spontaneous breaking of $B-L$ and R-parity 
at the soft supersymmetry breaking scale and furthermore dictates the presence of two sub-eV sterile neutrinos. See Refs.~\cite{Collider,Neutrinos1,Neutrinos2} for a detailed study of this model. 
See also Refs.~\cite{paper1,paper2,paper3,paper4,paper5,paper6} for other ideas related to R-parity violation.

In this article we investigate in details the most striking possible signals at the Large Hadron Collider: the lepton number violating signal of four leptons, three of them with the same sign, and four jets at the 7,8 and 14 TeV LHC. These channels result from pair produced charged sleptons (not currently tightly constrained by LHC data) which subsequently decay into neutralinos LSPs and provide an unique smoking gun for lepton number and RPV at the LHC. We further investigate the thermalization of the two light sterile neutrinos through the $B-L$ gauge boson, $Z_{BL}$ in the early universe showing that this model is consistent with the recent Planck results.

This article is organized as follows: In Section II we review the minimal gauge theory for R-parity. In Section III we investigate in detail 
the leptonic signals from RPV, while in Section IV we discuss the bounds coming from cosmology. Our main results 
are summarized in Section V.
%%%%%%%%%%%%%%%%%%%%%%%%%%%%%%%%%%%%%%%%%%%%%%%%%%%%%%%%%%%%%%%%%%%%%%%
\section{The Theory of R-parity}
%%%%%%%%%%%%%%%%%%%%%%%%%%%%%%%%%%%%%%%%%%%%%%%%%%%%%%%%%%%%%%%%%%%%%%%
We have shown in Ref.~\cite{PRL} that the simplest theory where one can understand dynamically the origin of RPV corresponds 
to the minimal supersymmetric $B-L$ theory based on
\begin{displaymath}
SU(3)_C \otimes SU(2)_L \otimes U(1)_Y \otimes U(1)_{B-L}.
\end{displaymath} 
In this context the matter content is given by
\begin{displaymath}
\hat{Q} \sim (3,2,1/6,1/3), \  \,  \hat{u}^c \sim (\bar{3},1,-2/3,-1/3), 
\end{displaymath}
\begin{displaymath}
\hat{d}^c \sim (\bar{3},1,1/3,-1/3), \  \ \hat{L} \sim (1,2,-1/2,-1),
\end{displaymath}
\begin{displaymath}
\hat{e}^c \sim (1,1,1,1), \  \,   {\rm{and}} \  \  \hat{\nu}^c \sim (1,1,0,1), \\
\end{displaymath}
where the presence of the right-handed neutrino is required by anomaly cancellation. The Higgs sector is composed of the MSSM Higgses
\begin{displaymath}
\hat{H}_u \sim (1,2,1/2,0), \  \text{and} \ \  \hat{H}_d \sim (1,2,-1/2,0).
\end{displaymath}
The superpotential of this theory is quite simply given by 
\begin{equation}
{\cal{W}}_{B-L}= Y_u \hat{Q} \hat{H}_u \hat{u}^c - Y_d \hat{Q} \hat{H}_d \hat{d}^c - Y_e \hat{L} \hat{H}_d \hat{e}^c + Y_\nu \hat{L} \hat{H}_u \hat{\nu}^c.
\end{equation}
As shown in Ref.~\cite{PRL}, the local B-L symmetry can be broken by the vacuum expectation value (VEV) of the right-handed sneutrino, without the addition of new $B-L$ Higgs fields. The VEV of the right-handed sneutrino further induces a VEV in the left-handed sneutrinos. Defining $\left< \tilde \nu^c_3 \right> \equiv v_R/\sqrt 2$, $\left< \tilde \nu_i \right> \equiv n_i/\sqrt 2$, $\left< \tilde H_u^0 \right> \equiv v_u/\sqrt 2$ and $\left< \tilde H_d^0 \right> \equiv v_d/\sqrt 2$, the sneutrino VEVs are given by
\begin{align}
	v_R & = \sqrt{\frac{-8 M_{\tilde \nu^c}^2}{g_{BL}^2}},
	\\
	n_i &= \frac{1}{\sqrt 2} \frac{v_R \left( {Y_\nu}_{i3} \, \mu \, v_d - {a_\nu}_{i3} v_u \right)}
		{M_{\tilde L_i}^2 - \frac{1}{2} M_{Z_{BL}}^2 + \frac{1}{2} \cos 2 \beta M_Z^2},
\end{align}
where $M_{Z_{BL}}^2 = g_{BL}^2 v_R^2/4$, and $a_\nu$ is the soft trilinear analogue of $Y_\nu$. $M_{\tilde \nu^c}^2$ and $M_{\tilde L}^2$ 
are the soft masses for the right-handed sneutrino and left-handed slepton doublet, respectively. Note that the first must be 
tachyonic for successful symmetry breaking. This has been discussed in the context of radiative 
symmetry breaking in~\cite{Ambroso:2009jd}. Furthermore, the breaking of $B-L$ induces a contribution to the soft scalar masses due to the D-terms:
\begin{equation}
	\Delta m_\phi^2 = \frac{1}{2} Q^{BL}_\phi M_{Z_{BL}}^2,
\end{equation}
where $Q^{BL}_\phi$ is the $B-L$ charge of $\phi$. One can quickly show that the left-handed sleptons receive the most negative contribution to their soft masses and we use this as a motivation to consider their accessibility at the LHC in the later sections.

Due to it's odd R-parity charge, the acquisition of a VEV by the right-handed sneutrino leads to spontaneous RPV which mixes same-spin same-electric charge particles together: neutrinos with neutralinos, charginos with charged leptons, Higgs bosons with sneutrinos and charged Higgs bosons with charged sleptons. Typically, the most important of these is the so-called bilinear term:
\begin{equation}
	W \supset \frac{1}{\sqrt 2} {Y_\nu}_{i3} v_R \hat{L}_i \hat{H}_u.
\end{equation}
The induced left-handed sneutrino VEV also leads to the following bilinear terms:
\begin{equation}
	\mathcal{L} \supset - \frac{1}{2}n_i \left[ g_2 \left(\sqrt 2 \, e_i \tilde W^+ +  \nu_i \tilde W^0\right) - g_1 \nu_i \tilde B - g_{BL} \nu_i \tilde B_{BL} \right].
\end{equation}
Since such bilinear terms lead to tree-level neutrino masses, their size is constrained to the level at which they can only affect the decays of the LSP, see Table~\ref{tbl.decays} for possible decays, leaving all other SUSY collider phenomenology unchanged. 

This simple supersymmetric theory predicts that
\begin{itemize}

\item R-parity must be broken leading to lepton number violation processes at the LHC.

\item The baryon number violating operators are generated at the non-renormalizable level, therefore satisfying the proton decay limits.

\item Two sub-eV sterile neutrinos must be present.

\item The $B-L$ and R-parity violating scales are defined by the soft masses of the right-handed sneutrinos, i.e. the soft SUSY scale.

\end{itemize} 
Furthermore, it is important to note that despite the violation of R-parity, an LSP gravitino can live long enough to potentially play the role of dark matter~\cite{Takayama:2000uz,Buchmuller:2007ui}. The phenomenological aspects  of this simple theory have been discussed in Ref.~\cite{Collider}. 
See also Refs.~\cite{Neutrinos1,Neutrinos2} for the properties of the neutrino sector.
%%%%%%%%%%%%%%%%%%%%%%%%%%%%%%%%%%%%
\section{Signatures at the LHC}
%%%%%%%%%%%%%%%%%%%%%%%%%%%%%%%%%%%%
We have mentioned in the previous section that the minimal supersymmetric $B-L$ theory predicts lepton number violation at the LHC 
and the relevant couplings are small. Therefore, the production mechanisms for the supersymmetric particles at the LHC are 
not modified but the LSP is not stable and it's decays via the lepton number violating couplings. In this case we could have 
the productions of several SM particles together with two LSPs
\begin{displaymath}
p \ p \  \to \  \Psi_1 \ldots \Psi_n \ {\rm{LSP}} \ {\rm{LSP}} \  \to \   \Psi_1 \ldots \Psi_n \  \Psi_i \ \Psi_j,
\end{displaymath}
where $\Psi_i$ is a SM particle, and $n=0,2,4,6,..$. One example is the scenario where the stop is the LSP and decays as a leptoquark:
\begin{displaymath}
p \ p \  \to \  \tilde{t}_1^* \  \tilde{t}_1 \to \   b \ \bar{b} \ \tau^+ \ \tau^-.
\end{displaymath}
In Table~\ref{tbl.decays} we show the possible LSP scenarios and their main decays, focusing only on light third generation sfermions.
\begin{table}[h]
\begin{tabular}{|c|c|c|c|}
\hline
~~~~~~~~ LSP  Scenario~~~~~~~~&~~~~ Decays ~~~~\\ \hline \hline
$\tilde{t}_1$ & $t \ \bar{\nu}, \  j \ \bar{\nu}, \ b \ e^+_i, \ j \ e^+_i$ \\ \hline
$\tilde{b}_1$ & $b \ \bar{\nu}, \  j \ \bar{\nu}, \ t \ e^-_i, \ j \ e^-_i$ \\ \hline
$\tilde{\chi}^0_1$ & $ e^{\pm}_i \ W, \  \nu \  Z, \  e^{\pm}_i \ H^{\mp}, \  \nu \ H_i^0$ \\ \hline
$\tilde{\chi}^{\pm}_1$ & $ e^{\pm}_i \ Z, \  \nu \  W^\pm, \  e^{\pm}_i \ H_i^0, \  \nu \ H_i^0$ \\ \hline
$\tilde{\tau}^{\pm}$ & $ e^{\pm}_i \nu, \  \bar{q} \ q', \ h W^\pm $ \\ \hline
$\tilde{\nu}_3$ & $ \bar{q} q, \ \bar{e}_i e_j, \ WW, \ ZZ, hh, HH $ \\ \hline
\end{tabular}
\caption{Lepton number violating LSP decays.}
\label{tbl.decays}
\end{table}
As discussed in Ref.~\cite{Collider}, the most exotic signals in this theory correspond to a neutralino LSP. 
In this case we can produce two charged sleptons through the photon, the $Z$ and the $B-L$ gauge boson:
\begin{equation}
p \ p \  \to \ \gamma, Z, Z_{B-L} \  \to \ \tilde{e}_i^+ \tilde{e}_i^-.  
\end{equation}  
The sleptons subsequently decay as $\tilde{e}_i  \to e_i \  \tilde{\chi}^0_1$ and finally the neutralinos decay through $\tilde{\chi}^0_1 \to e^{\pm}_j W^{\mp}$. In Ref.~\cite{Collider}, it was shown shown that both of these branching ratios are typically large. Therefore, one can expect a significant number of events with four leptons and two W's.
When the Ws decay hadronically, one is left with the unambiguous lepton number violating final states:
\begin{eqnarray}
e_i^{\pm} \ e^{\mp}_i \ e^{\pm}_j \ e^{\pm}_k \ 4 j.
\end{eqnarray}
where $e_i=e,\mu,\tau$. The number of events for these channels with multileptons is given by
\begin{equation}
N_{ijk}={\cal{L}} \times \sigma (p p \to \tilde{e}^{\pm}_i \tilde{e}^\mp_i) \times C_{jk},
\end{equation}
where ${\cal{L}}$ is the integrated luminosity, and $C_{jk}$ is given by
\begin{equation}
C_{jk}= 2 (2-\delta_{jk}) \times {\rm{Br}} (\tilde{e}^{\pm} \to e^{\pm} \tilde{\chi}_1^0)^2 \times {\rm{Br}} (\tilde{\chi}^0_1 \to e^{\pm}_j W^{\mp}) \times {\rm{Br}} (\tilde{\chi}^0_1 \to e^{\pm}_k W^{\mp}) 
\times {\rm{Br}} (W \to jj)^2. 
\end{equation} 
In the next section we investigate the predictions for these channels in different scenarios.
%
%%%%%%%%%%%%%%%%%%%%%%
\subsection{Slepton Production at the LHC}
\begin{figure}[h] 
	\includegraphics[scale=0.9]{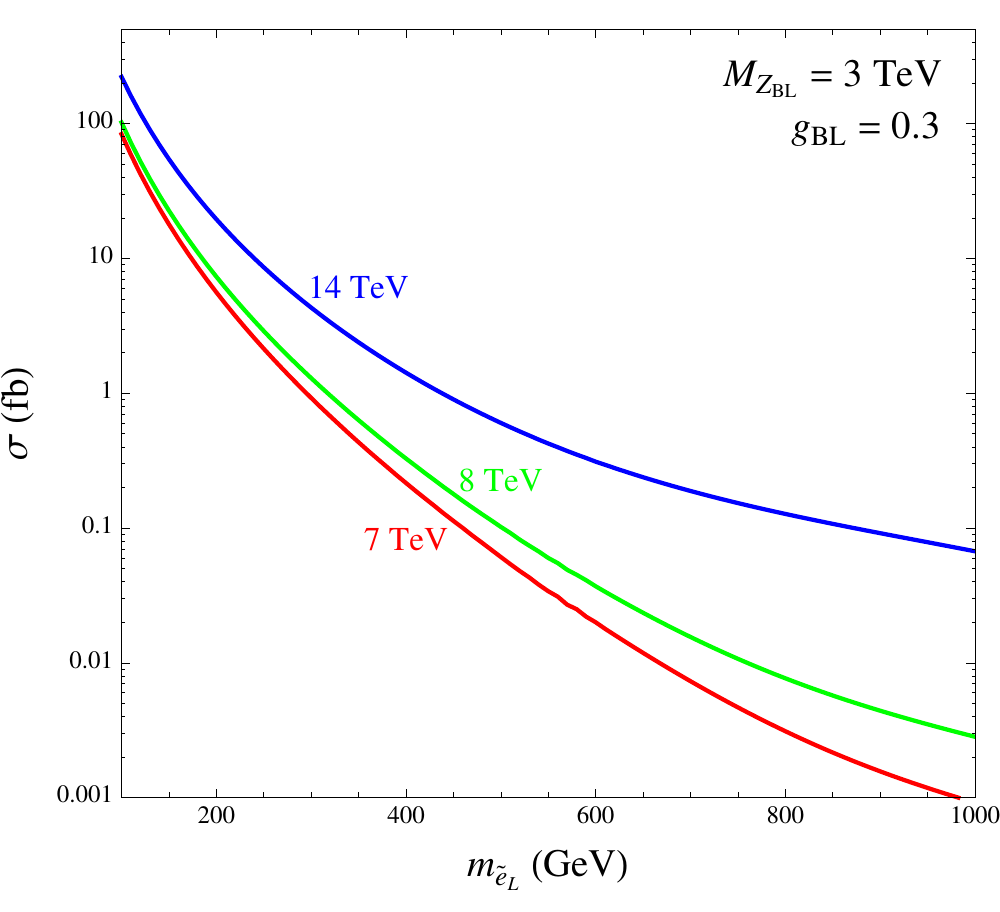}
	\caption{Cross sections for $p\, p \to \gamma, Z,Z_{BL} \to \tilde{e}^{\pm}_i \tilde{e}^{\mp}_i$ with LHC center of mass energies of 7 TeV (red), 8 TeV (green) and 14 TeV (blue), assuming $M_{Z_{BL}} = 3$ TeV and $g_{BL}=0.3$ versus the left-handed slepton mass.}
\label{cs}
\end{figure}

%%%%%%%%%%%%%%%%%%%%%%%
In order to analyze the testability of the channels with multi-leptons we need to estimate the cross section using the production mechanism mentioned above and focusing on left-handed sleptons\footnote{We use the notation $\tilde e_i$ to indicate the $i^\text{th}$ generation left-handed slepton.}, as motivated above. In Fig.~\ref{cs} we show 
the production cross section for left-handed sleptons assuming that the mass of the $B-L$ gauge boson is $M_{Z_{BL}}=3$ TeV and the corresponding coupling 
is $g_{BL}=0.3$. In this figure we show the numerical results for $\sigma (p p \to \tilde{e}^{\pm}_i \tilde{e}^\mp_i)$ when the center mass energy is $\sqrt{s}=7$ TeV,  
$\sqrt{s}=8$ TeV, and $\sqrt{s}=14$ TeV. It is important to mention that when $\sqrt{s}=14$ TeV the cross section is above 1 fb when the slectron mass is below 400 GeV. For selectron masses in the TeV range, the signals discussed in this paper will be very difficult to test.
%
%%%%%%%%%%%%%%%%%%%%%%%%%%
\subsection{Multi-Lepton Channels}
\begin{figure}[t!] 
	\includegraphics[scale=0.9]{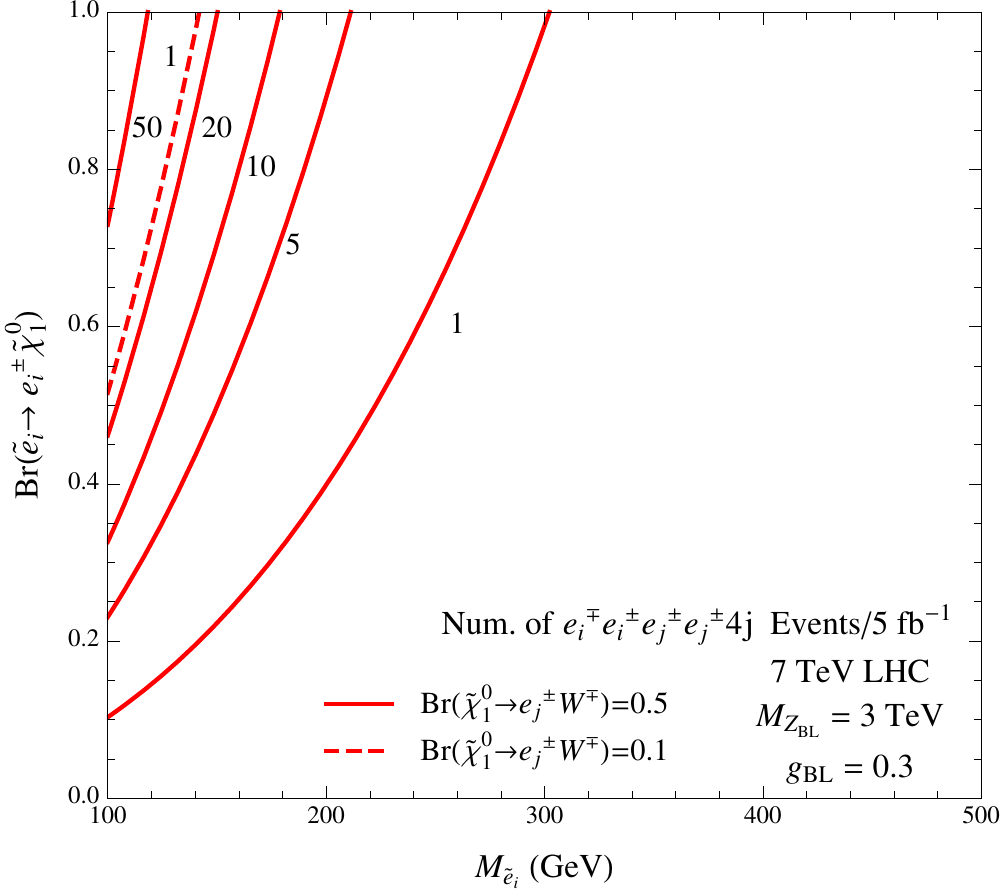}
	%\put(-104,-4){(a)}
	\includegraphics[scale=0.9]{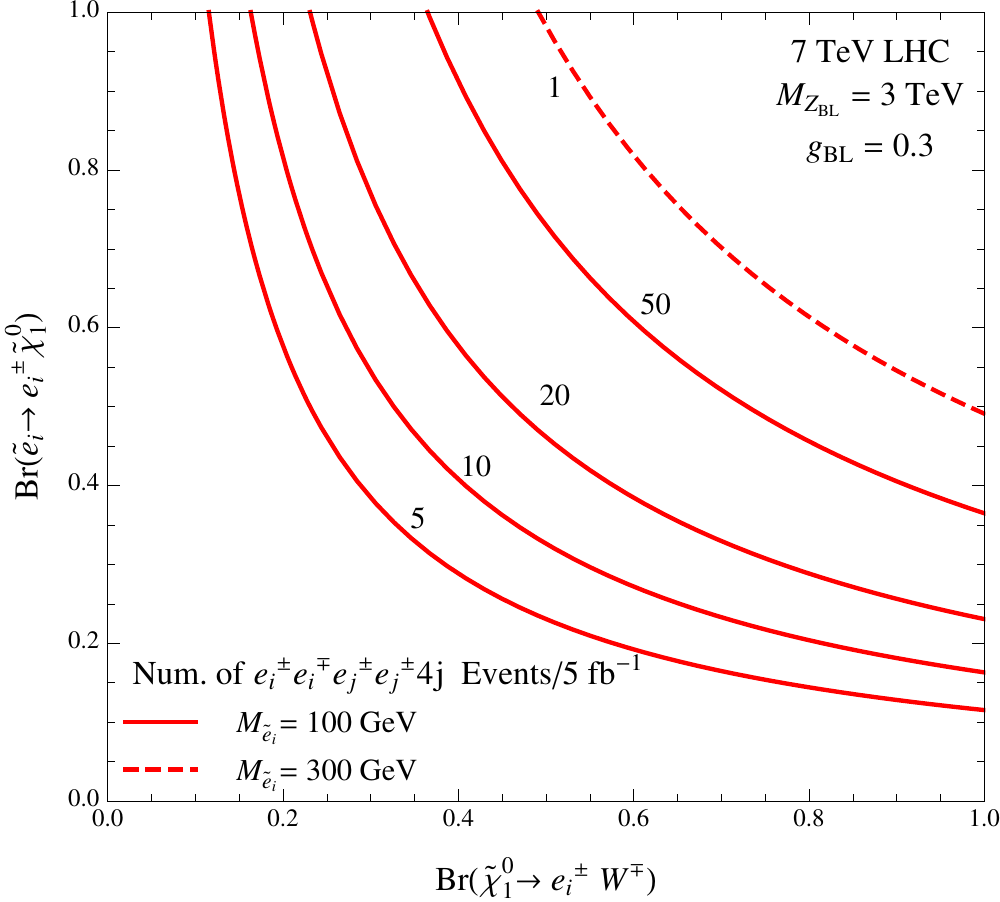}
	%\put(-104,-4){(b)}
	\caption{Number of events in the ${\rm{Br}} (\tilde{e}^{\pm}_i \to e^{\pm}_i \tilde{\chi}_1^0)$ vs $M_{\tilde{e}_i}$ plane in the upper panel assuming $\sqrt{s}=7$ TeV, ${\rm{Br}} (\tilde{\chi}^0_1 \to e^{\pm}_k W^{\mp})=0.1 (0.5)$, 
when $M_{Z_{BL}}=3$ TeV and ${\cal{L}}=5 \ {\rm{fb}}^{-1}$. In the lower panel the number of events are plotted in the plane 
${\rm{Br}} (\tilde{e}^{\pm}_i \to e^{\pm}_i \tilde{\chi}_1^0)$-${\rm{Br}} (\tilde{\chi}^0_1 \to e^{\pm}_k W^{\mp})$.
}
\label{7TeV}
\end{figure}

\begin{figure} [t!]	
	\includegraphics[scale=0.9]{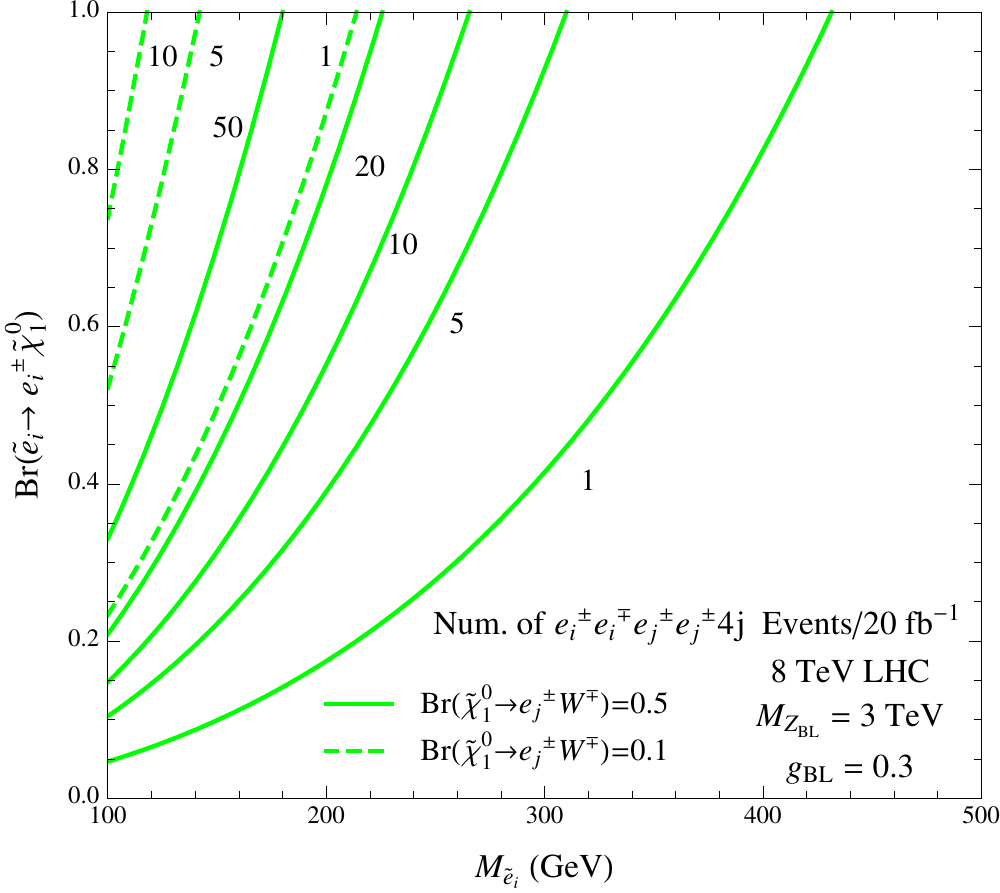}
	%\put(-104,-4){(c)}
	\includegraphics[scale=0.9]{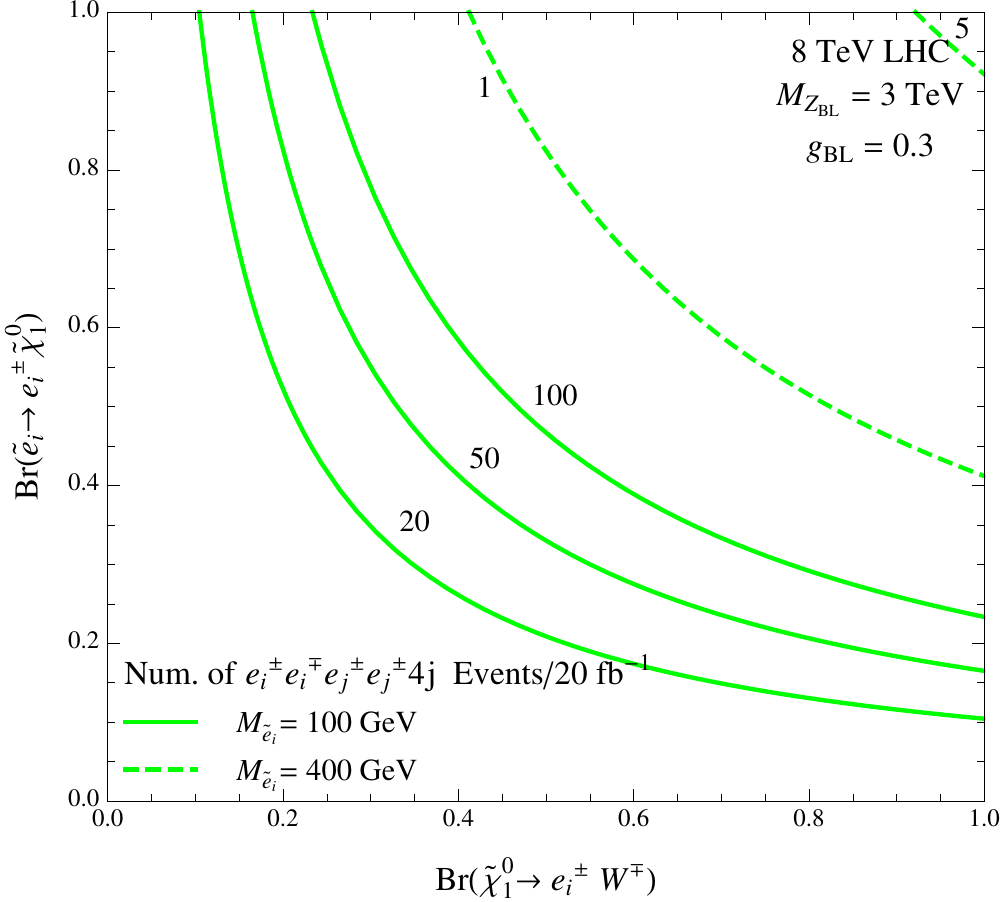}
	%\put(-104,-4){(d)}
	\caption{Number of events in the ${\rm{Br}} (\tilde{e}^{\pm}_i \to e^{\pm}_i \tilde{\chi}_1^0)$ vs $M_{\tilde{e}_i}$ plane in the upper panel assuming 
	$\sqrt{s}=8$ TeV, ${\rm{Br}} (\tilde{\chi}^0_1 \to e^{\pm}_k W^{\mp})=0.1 (0.5)$, when $M_{Z_{BL}}=3$ TeV and ${\cal{L}}=20 \ {\rm{fb}}^{-1}$. 
	In the lower panel the number of events are plotted in the plane ${\rm{Br}} (\tilde{e}^{\pm}_i \to e^{\pm}_i \tilde{\chi}_1^0)$-${\rm{Br}} (\tilde{\chi}^0_1 \to e^{\pm}_k W^{\mp})$.}
	\label{8TeV}
\end{figure}

\begin{figure}[h!]
	\includegraphics[scale=0.9]{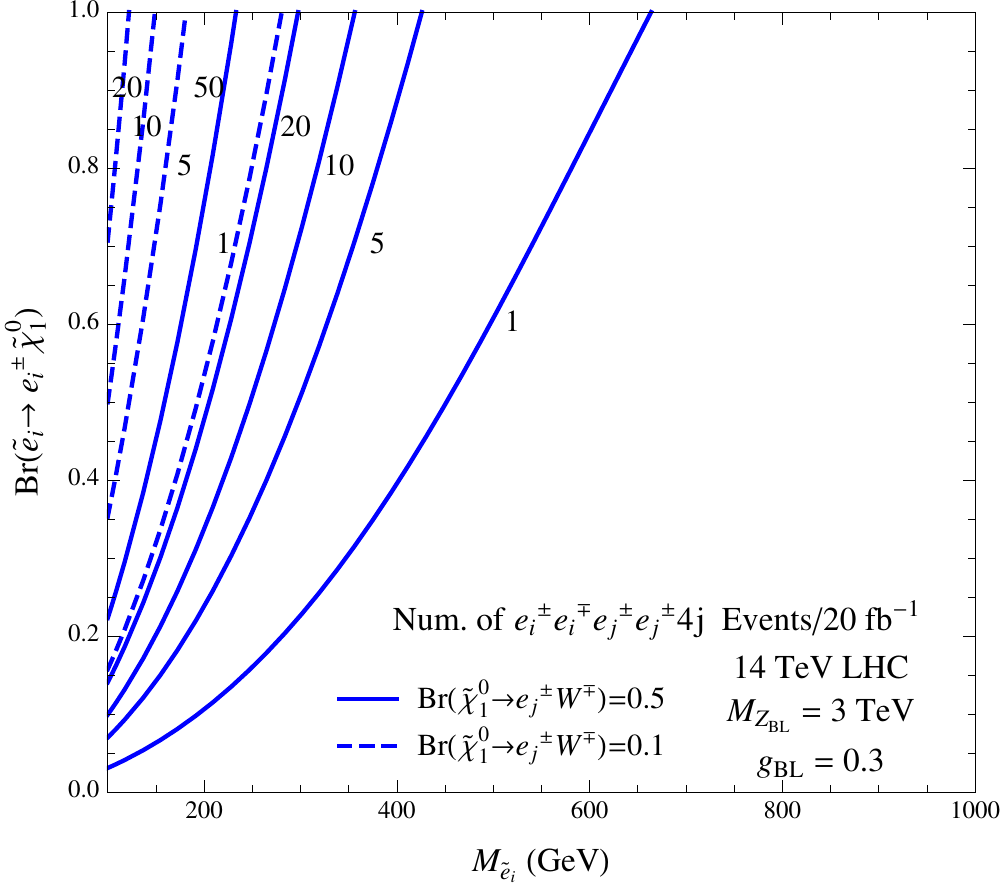}
	%\put(-104,-4){(e)}
	\includegraphics[scale=0.9]{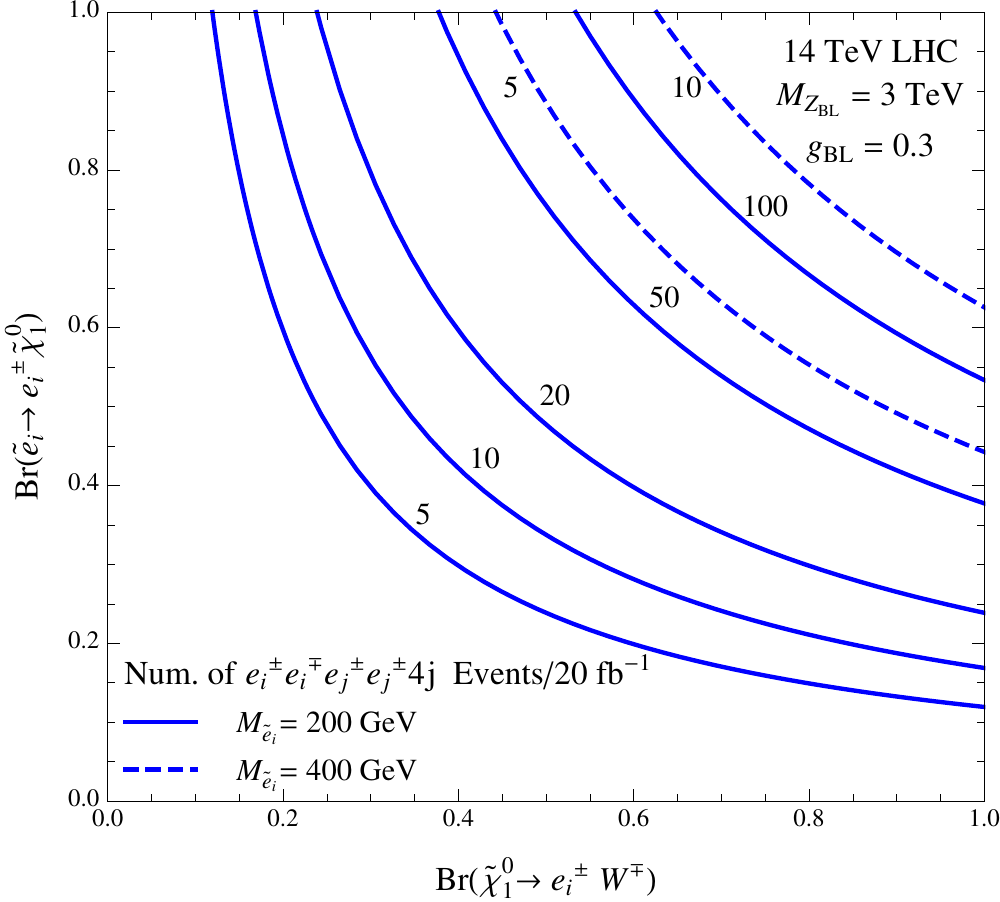}
	%\put(-104,-4){(f)}	
	\caption{Number of events in the ${\rm{Br}} (\tilde{e}^{\pm}_i \to e^{\pm}_i \tilde{\chi}_1^0)$ vs $M_{\tilde{e}_i}$ plane in the upper panel assuming 
	$\sqrt{s}=14$ TeV, ${\rm{Br}} (\tilde{\chi}^0_1 \to e^{\pm}_k W^{\mp})=0.1 (0.5)$, when $M_{Z_{BL}}=3$ TeV and ${\cal{L}}=20 \ {\rm{fb}}^{-1}$. 
	In the lower panel the number of events are plotted in the plane ${\rm{Br}} (\tilde{e}^{\pm}_i \to e^{\pm}_i \tilde{\chi}_1^0)$-${\rm{Br}} (\tilde{\chi}^0_1 \to e^{\pm}_k W^{\mp})$.}
\label{14TeV}
\end{figure}

%%%%%%%%%%%%%%%%%%%%%%%%%%
%
The number of events with four leptons and four jets are estimated in three different scenarios:
$\sqrt{s}=7$ TeV and 5 fb$^{-1}$ in Fig.~\ref{7TeV}, $\sqrt{s}=8$ TeV and 20 fb$^{-1}$ in Fig.~\ref{8TeV} and $\sqrt{s}=14$ TeV and 20 fb$^{-1}$ in Fig.~\ref{14TeV}. The upper panels in each of these figures shows the number of events in the ${\rm{Br}} (\tilde{e}^{\pm}_i \to e^{\pm}_i \tilde{\chi}_1^0)$--$M_{\tilde{e}_i}$ plane assuming an optimistic (pessimistic) branching ratio for $\tilde \chi^0_1 \to e_i^\pm W^\mp$ of 0.5 (0.1). In the lower panels, the number of events is shown in the ${\rm{Br}} (\tilde{e}^{\pm}_i \to e^{\pm}_i \tilde{\chi}_1^0)$--$M_{\tilde{e}_i}$ for a light (heavy) selectron mass of 100 GeV (300 GeV) for the 7 TeV run, 100 GeV (400 GeV) for the 8 TeV run and 200 GeV (400 GeV) for the 14 TeV run. In all figures $M_{Z_{BL}}=3$ TeV and $g_{BL}=0.3$ has been assumed.

In general, of course, these figures show that the number of events is maximized for small slepton masses and large branching ratios of slepton into lepton neutralino and neutralino into lepton $W$. For the 7 TeV run a 300 GeV selectron is already too heavy to be observable while at 14 TeV and 20 fb$^{-1}$ one can potentially hope to observe sleptons as heavy as 500 GeV.

Some simple benchmark scenarios are defined in Table~\ref{tbl.num.events} to establish a sense of the number of events expected for the 14 TeV run after 80 fb$^{-1}$ of data. The benchmarks are for a 400 GeV and 600 GeV slepton assuming $\text{Br}\left(\tilde e_i \to e_i \tilde \chi^0_1\right) = 1$ and $\text{Br}\left(\tilde \chi_1^0 \to e^\pm W^\mp\right) = 0.2$ while $\text{Br}\left(\tilde \chi_1^0 \to \mu^\pm W^\mp\right) = 0.4$. This scenario is motivated by a normal hierarchy in the neutrino sector which translates into typically lower branching ratios into electrons. Furthermore, while the masses are quite heavy they were chosen so that the resulting leptons have $p_T$ high enough to pass selection cuts. As in the Figures, it is assumed that  $M_{Z_{BL}} = 3$ TeV and $g_{BL}=0.3$. As the LHC is expected to surpass 80 fb$^{-1}$, it is possible that one can probe even beyond 600 GeV charged sleptons. It also important to remember the resonance behavior as the slepton mass approaches half the $Z_{BL}$ mass which would prevent the cross section from dropping rapidly.
\begin{table}[htdp]
\caption{Number of events for the channels with multileptons for $\sqrt{s}=14$ TeV and ${\cal{L}}=80 \ {\rm{fb}}^{-1}$, assuming a branching ratio $\tilde e_i \to e_i \tilde \chi^0_1$ of one and $\text{Br}\left(\tilde \chi_1^0 \to e^\pm W^\mp\right) = 0.2$ while $\text{Br}\left(\tilde \chi_1^0 \to \mu^\pm W^\mp\right) = 0.4$.}
\begin{center}
\begin{tabular}{|c|c|c|c|}
\hline
 $M_{\tilde{e}_i}=400$ GeV & $\sigma (p p \to \tilde{e}^{\pm}_i \tilde{e}^\mp_i)= 1.4$ fb  & & \\
\hline 
 & Channels  & Combinational Factor  & Number of Events \\
 \hline
  & $e_i^{\pm} e_i^{\mp}  \mu^{\pm} \mu^{\pm}  4 j$ &0.15 & 16\\
   \hline
  & $e_i^{\pm} e_i^{\mp}  \mu^{\pm} e^{\pm}  4 j$ & 0.15& 16\\
  \hline
  & $e_i^{\pm} e_i^{\mp}  e^{\pm} e^{\pm}  4 j$ & 0.037& 4\\
\hline
 $M_{\tilde{e}_i}=600$ GeV & $\sigma (p p \to \tilde{e}^{\pm}_i \tilde{e}^\mp_i)= 0.31$ fb&& \\
 \hline
 & Channels & Combinational Factor & Number of Events  \\
 \hline
  & $e_i^{\pm} e_i^{\mp}  \mu^{\pm} \mu^{\pm}  4 j$ & 0.15 & 4 \\
   \hline
  & $e_i^{\pm} e_i^{\mp}  \mu^{\pm} e^{\pm}  4 j$ & 0.15 & 4 \\ 
   \hline
  & $e_i^{\pm} e_i^{\mp}  e^{\pm} e^{\pm}  4 j$ & 0.037 & 1 \\  
 \hline
\end{tabular}
\end{center}
\label{tbl.num.events}
\end{table}%
%\newpage
%%%%%%%%%%%%%%%%%%%%%%%%%%%%%%
\section{Constraints from Cosmology}
%%%%%%%%%%%%%%%%%%%%%%%%%%%%%%
The presence of sub-eV sterile neutrinos predicted by the minimal theory of R-parity have cosmological implications as they contribute to the radiation content of the universe. Such dark radiation is parameterized as the number of effective thermalized neutrino species, $N_\text{eff}$, and impacts several cosmological events including nucleosynthesis and the time of matter-radiation equality. These in turn affect the element abundance today, the cosmic microwave background (CMB) and the value of Hubble's constant ($H_0$). The most recent Planck results constrain $N_\text{eff}$ as follows~\cite{Ade:2013zuv}:
\begin{align}
\begin{split}
&	3.36 \pm 0.34 : \quad \text{CMB only},
	\\
&	3.62 \pm 0.25 : \quad \text{CMB}+ H_0,
\end{split}
\end{align}
where the three active neutrinos contribute a value of $N_\text{eff} = 3.045$.

The contribution of the sterile neutrinos to $N_\text{eff}$ depends on the degree to which they have been thermalized. Thermalization can take place through two mechanisms: sterile-active oscillations and/or mediation via new gauge bosons, in our case a neutral gauge boson $Z_{BL}$. The former was first considered in works such as \cite{Barbieri:1989ti, Enqvist:1990dq}, and more recently in \cite{Hannestad:2012ky, Mirizzi:2012we}. These later studies have shown that cosmological data is in tension with sterile-active mixings necessary to explain terrestrial neutrino anomalies, unless large lepton asymmetries are assumed. However, it is important to keep in mind though that the sterile neutrinos discussed here have not been introduced to confront the short baseline anomalies but are a consequence of the model. Therefore, it is possible that their mixings with the active neutrinos are quite small (as assumed in the collider study) and therefore these cosmological bounds from sterile-active mixings would not play a role here. Alternatively, one can study this issue in the presence of lepton asymmetries.

Leaving this issue aside, we confront the latter mechanism of sterile neutrino thermalization: mediation via $Z_{BL}$. An analysis along these lines places bounds on the ratio $M_{Z_{BL}}/g_{BL}$ which can be relevant for the LHC. Such a study has been conducted for several $U(1)$ extensions of the SM gauge group motivated by GUTs~\cite{SolagurenBeascoa:2012cz,Anchordoqui:2012qu} and we follow the analysis of the former reference here.

To begin, one must determine the decoupling temperature of the sterile neutrinos, $T_\text{dec}^{\nu_R}$, which occurs when the Hubble parameter overtakes the rate of the reaction which keeps the sterile neutrinos in equilibrium, in this case annihilation into SM particles through $Z_{BL}$. The rate is calculated as
\begin{eqnarray}
	\Gamma(T) &=& n_{\nu_R} (T) \left< \sigma (\bar{\nu}_R \nu_R \  \to {\rm{SM}} \  {\rm{SM}} ) v \right> \\
	&=& \sum_{f} \frac{g_{\nu_R}^2}{8 \pi^4 n_{\nu_R}(T)}
		\int_0^\infty p^2 dp
		\int_0^\infty q^2 dq
		\int_{-1}^1 \frac{1 - \cos \theta }{\left(e^{q/T} + 1\right)\left(e^{p/T} + 1\right)} \sigma_f(s) d(\cos \theta) ,
\end{eqnarray}
where
\begin{equation}
n_{\nu_R} (T)= \frac{g_{\nu_R}}{2 \pi^2} \int_0^\infty p^2 dp  \frac{1}{\left(e^{p/T} + 1\right)},
\end{equation}
and
\begin{equation}
	\sigma_f(s) = \frac{N_C^f {(Q^{BL}_f)}^2 {(Q^{BL}_{\nu_R})}^2}{12 \pi} \frac{g_{BL}^4}{M_{Z_{BL}}^4} s
				\left(1 + \frac{2 m_f^2}{s} \right)
				\sqrt{1- \frac{4 m_f^2}{s}}.
\end{equation}
Here, $g_{\nu_R} =2$ is the number of right-handed neutrino degrees of freedom for one generation, the exponential factors are the Fermi-Dirac distributions, while $s = 2pq\left( 1 - \cos \theta \right)$ is the center of mass energy, $v = \left(1 - \cos \theta \right)$ is the relative velocity and $\theta$ is the relative angle of the colliding sterile neutrinos. The Hubble parameter is given by
\begin{equation}
	H(T) = 1.66 \sqrt{g(T) + \frac{7}{2} \ }  \times \frac{T^2}{M_{Pl}},
\end{equation}
where $g(T)$ is the number of relativistic SM species in thermal equilibrium at temperature T. This can be expressed as 
\begin{equation}
g(T) = g_B(T) + \frac{7}{8} g_F(T).
\end{equation}
Here $g_{B,F}$ are the number of relativistic degrees of freedom for bosons and fermions in thermal equilibrium, respectively. The addition of 7/2 to g(T) is the contribution of the two sterile neutrinos predicted by the theory for R-parity. The calculation of $g(T)$ must be carried out using lattice techniques due to the presence of the QCD phase transition at temperatures around 155 MeV. To this end, the results of~\cite{Laine:2006cp} are adopted here.

\begin{figure}[h]
\includegraphics[scale=0.7]{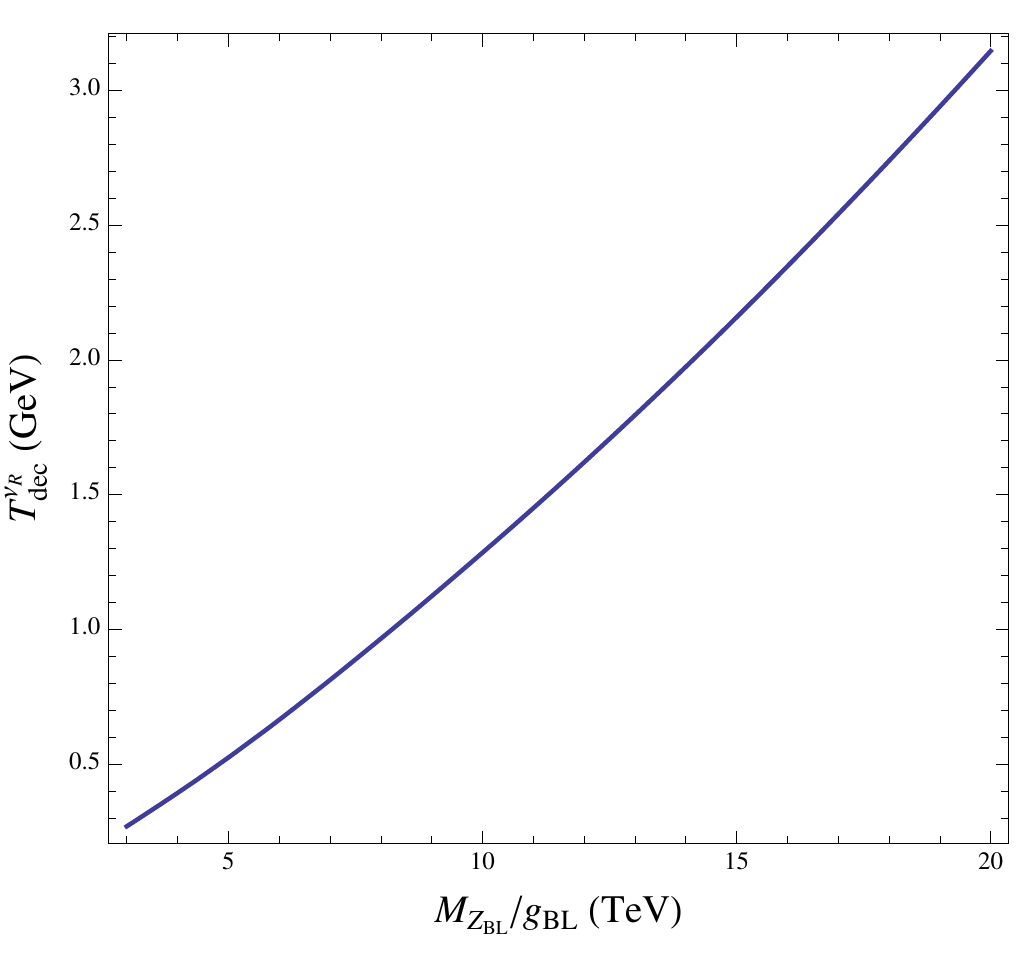}
%\put(-104,-4){(e)}
\includegraphics[scale=0.7]{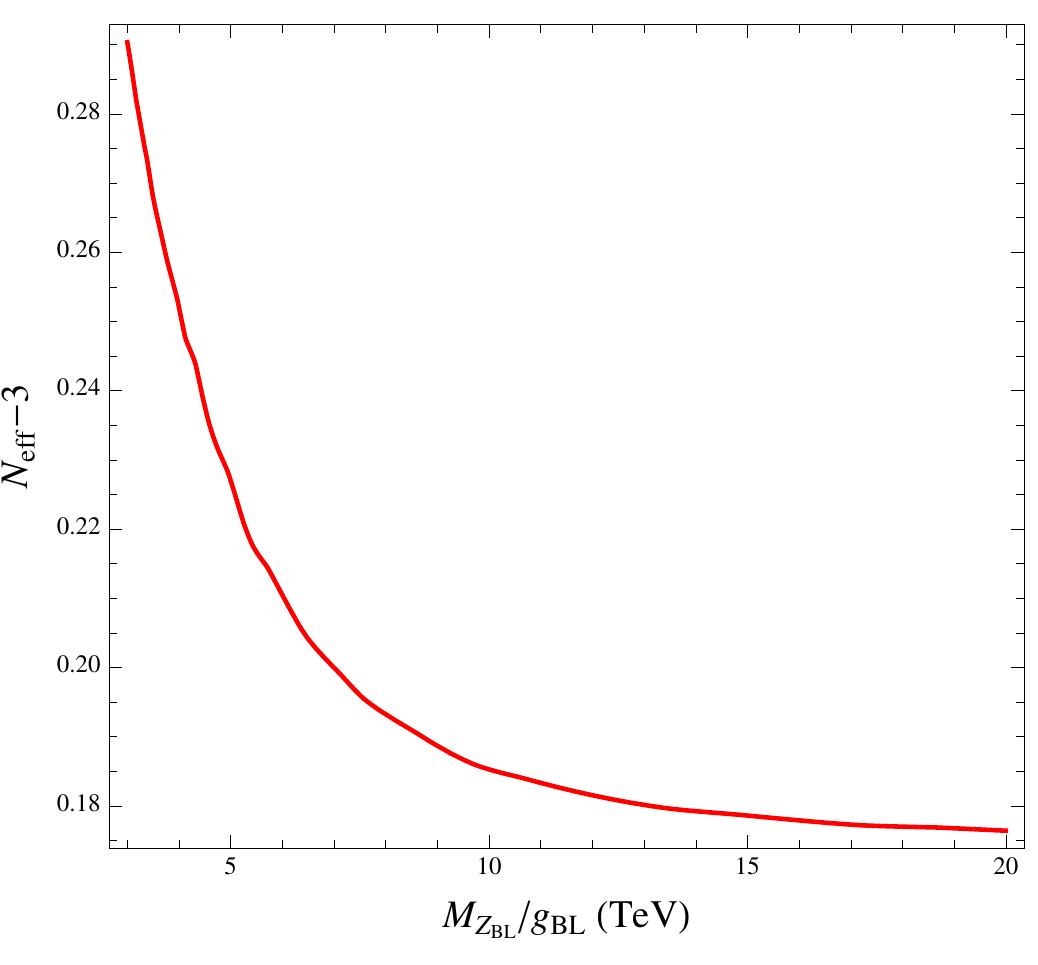}
%\put(-104,-4){(f)}	
\caption{Left panel: the decoupling temperature of the sterile neutrinos versus the ratio of the gauge boson 
mass to the gauge coupling. Right panel: number of effective neutrinos versus the ratio of the gauge boson mass 
to the gauge coupling.}
\label{cosmo}
\end{figure}

Using $H(T_\text{dec}^{\nu_R}) = \Gamma(T_\text{dec}^{\nu_R})$, one can solve for $T_\text{dec}^{\nu_R}$ numerically yielding the results plotted in the left-handed 
side of Fig.~\ref{cosmo} versus the ratio $M_{Z_{BL}}/g_{BL}$. The benchmark point discussed in the collider section corresponds to a ratio of 10 TeV and therefore a 
decoupling temperature of 1.3 GeV. Finally, for two sterile neutrinos and the three active neutrinos, $N_\text{eff}$ is given by
\begin{equation}
	N_\text{eff} = 3.045+ 2 \left(\frac{g(T_d^{\nu_L})}{g(T_d^{\nu_R})}\right)^{\frac{4}{3}},
\end{equation}
where $g(T_d^{\nu_L}) = 43/4$. The results are plotted in the right-hand side of Fig.~\ref{cosmo}. This range is using the central values for the Planck one sigma data. However, it is important to remember that it is the lowest possible value of $N_\text{eff}$ in this model (without a significant lepton asymmetry), since significant sterile-active oscillations would increase it. For the benchmark used in the collider study, $N_\text{eff} = 3.2$, which is in agreement with cosmology.

These results show that in our model it is possible to have a consistent scenario in agreement with cosmology and collider physics.
%%%%%%%%%%%%%%%%%%%%%%%%%%%%%
\section{Concluding Remarks}
%%%%%%%%%%%%%%%%%%%%%%%%%%%%%
In this article we have investigated the most striking signals at the Large Hadron Collider 
in the context of the simplest gauge theory for R-parity. We have shown the numerical predictions 
for the pair production of sleptons at the LHC, and the number of events for the channels 
with four leptons and two jets. Here the existence of three leptons with the same electric 
charge is a smoking gun coming from the violation of the lepton number and the 
spontaneous breaking of R-parity.

In order to make sure that our results are in agreement with cosmology, we have 
investigated in section IV the constraints coming from the effective number of 
relativistic degree of freedom determined by the Planck collaboration. 
These constraints set a lower bound on the mass of the new $B-L$ gauge 
boson for a given value of the $B-L$ gauge coupling. All the scenarios discussed 
in this paper are in agreement with the bounds coming from collider physics 
and cosmology.

{\textit{Acknowledgment}}:
S. S. thanks the Particle and Astro-Particle Division at the Max Planck Institute for Nuclear Physics for hospitality.  

%%%%%%%%%%%%%%%%%%%%%%%%%%%%%
\appendix
%%%%%%%%%%%%%%%%%%%%%%%%%%%%%
%%%%%%%%%%%%%%%%%%%%%%%%%%%%%
\section{Feynman Rules and Cross Section}
%%%%%%%%%%%%%%%%%%%%%%%%%%%%%
The relevant Feynman rules for our study are given by
\begin{align}
	A \, \tilde e_L(p_3) \, \tilde e_L^*(p_4) & : - i e \, Q_e \left(p_4 - p_3 \right)_\mu,
	\\
	Z \, \tilde e_L(p_3) \, \tilde e_L^*(p_4) & : - i \frac{e}{s_W c_W} L_e \left(p_4 - p_3 \right)_\mu,
	\\
	Z_{BL} \, \tilde e_L(p_3) \, \tilde e_L^*(p_4) & : - i g_{BL}  Q_e^{(B-L)} \left(p_4 - p_3 \right)_\mu,
	\\
	A \, \bar{q} \, q & : - i e \, Q_q \gamma_\mu,
	\\
	Z \, \bar{q} \, q & : - i \frac{e}{s_W c_W} (L_q P_L + R_q P_R) \gamma_\mu,
	\\
	Z_{BL} \, \bar{q} \, q & : - i g_{BL} Q_q^{BL} \gamma_\mu,
\end{align}
where $s_x=\sin x$, $c_x = \cos x$. Here $P_L$, and $P_R$ are the left-handed and 
right-handed projection operators, respectively. $Q_f$ is the electric charge of 
fermion $f$, and $Q^{BL}_f$ is the half the $B-L$ charge of fermion $f$.
\begin{align}
	L_f & = I^3_f - Q_f s_w^2,
	\\
	R_f & = - Q_f s_w^2.
\end{align}
Here, $I^3_f$ is the left-handed isospin of fermion $f$. From these the differential cross section for slepton production at the LHC can be derived:
\begin{align}
\begin{split}
	\frac{d \sigma}{d t} = & \frac{1}{16 \pi} \frac{1}{s^2} |\bar{\mathcal M}|^2
	\\ 
	= & \frac{4 \pi \alpha^2}{3} \frac{\left(ut - m_{\tilde e_L}^4\right)}{s^2}
	\left[
		\frac{Q_q^2 \, Q_e^2}{ 2 s^2}
		+ \frac{Q_q Q_e \left(L_q + R_q\right) L_e}{2s_w^2 c_w^2 s \left(s-M_Z^2\right)}
		+ \frac{\left(L_q^2 + R_q^2\right)L_e^2}{4 s_w^4 c_w^4 \left(s-M_Z^2\right)^2}
		+ \frac{g_{BL}^2 Q_q Q_e Q^{BL}_q Q^{BL}_e}{4 \pi  \alpha s \left(s-M_{Z_BL}^2\right)}
	\right.
	\\
	& \left. \hspace{2.8cm}
		+ \frac{ g_{BL}^2 \left(L_q +R_q\right) L_e Q^{BL}_q Q^{BL}_e}{8 \pi \alpha s_w^2 c_w^2 \left(s-M_{Z}^2\right) \left(s-M_{Z_BL}^2\right)}
		+ \frac{g_{BL}^4 {Q^{BL}_q}^2 {Q^{BL}_e}^2}{32 \pi^2 \alpha^2 \left(\left(s-M_{Z_BL}^2\right)^2 + \Gamma_{Z_BL}^2 M_{Z_BL}^2\right)}
	\right].
\end{split}
\end{align}
%
%%%%%%%%%%%%%%%%%%%%%%%%%%%%%

\end{document}